\documentclass[final,5p,times,twocolumn]{elsarticle}
\newcommand{\pf}[2]{\frac{\partial ,1}{\partial ,2}}
\newcommand{\pfs}[2]{\frac{\partial^2 ,1}{\partial ,2 ^2}}
\newcommand{\pft}[2]{\frac{\partial^3 ,1}{\partial ,2 ^3}}
\newcommand{\pderiv}[2]{\frac{\partial ,1}{\partial ,2}}

\usepackage{multirow}
\usepackage{tabu}
\usepackage{CJKutf8}
\usepackage[T1]{fontenc}
\usepackage[utf8]{inputenc}
\usepackage{graphics, amsmath}
\usepackage{graphicx,color}
\usepackage{epsfig}
\usepackage{amsmath}
\usepackage{amssymb}
\usepackage{setspace}
\usepackage{hyperref}
\usepackage{cleveref}
\usepackage{underscore}
\usepackage[dvipsnames]{xcolor}
\usepackage{multirow, makecell}
\usepackage{colortbl}
\usepackage{dashrule}
\usepackage{ehhline}
\usepackage{csquotes}
\usepackage{natbib}
\biboptions{sort&compress}

\usepackage{array}
\usepackage{booktabs}
\usepackage{caption}
\makeatletter
\usepackage{colortbl}
\usepackage{hhline}

\begin{document}

\begin{frontmatter}
	\title{Implementation of nonlocal non-Fourier heat transfer for semiconductor nanostructures}
	\author[1]{Roya Baratifarimani}
\author[1]{Zahra Shomali\corref{cor1}}
\address[1]{Department of Physics, Basic Sciences Faculty, Tarbiat Modares University, Tehran, Iran}
%\address[2]{School of Physics, Institute for Research in Fundamental Sciences (IPM), Tehran 19395-5531, Iran}
	\cortext[cor1]{Corresponding author, Tel.: +98(21) 82883147.}
\ead{shomali@modares.ac.ir}

\begin{abstract}
	
The study of heat transport in micro/nanoscale structures due to their application, especially in Nanoelectronics, is a matter of interest. In other words, the precise simulation of the temperature distribution inside the transistors is consequential in designing and building more reliable devices reaching lower maximum temperatures during the operation. The present study constitutes a framework for micro/nanoscale heat transport study which leads to the calculation of accurate temperature/heat flux profiles with low computational cost. The newly non-dimensional parameter $\gamma$, presenting the strength of the nonlocality, is utilized through the nonlocal DPL modeling (NDPL). Alongside the calculating nonlocality coefficient, the factors also appearing in DPL, including the temperature jump, phase lagging ratio, are revisited. The factor $\gamma$ is found to have a linear relationship with Knudsen (Kn) number, being 3.5 for Kn=10 and 0.035 for Kn=0.1. Although the nonlocality is bold for the large Knudsen numbers, it also plays a vital role for low Knudsen number structures especially at earlier times. Further, It is obtained that intruding $\gamma$ is critical for obtaining accurate temperature and heat flux distributions which are very close to the practical results of Phonon Boltzmann equation.

\end{abstract}


\end{frontmatter}

%\linenumbers
%----------------------------------------------------------------------------------------------------
%\tableofcontents
\section{Introduction}

Micro/nanoscale heat conduction in nano-materials, especially the semiconductors, has attracted significant attention in recent years. This concern on one hand, relates to the practical application in nano-electric \cite{Mahajan2002,Moghadam2014}, and on the other hand, provides worthwhile understanding of the fundamental principles behind the nanoscopic heat carries' behavior \cite{Minnich2012,Shomali2019}. The Fourier's law, $\vec{q}(\vec{r},t)=-k\vec{\nabla}T(\vec{r},t)$, which is established as the accurate model to simulate heat conduction in macroscopic structures, fails when length/time scales are, respectively, comparable to the phonon mean free paths and relaxation times \cite{Chiu2005,Alvarez2007,Shomali2018}. 

Previously, many studies have been performed focusing on the heat transfer temporal behavior at the nanoscale. Specifically, the constitutive macroscopic equation introduced as the dual-phase-lag (DPL) has been suggested:
	
	 \begin{equation}
	 	\vec{q}(\vec{r},t+\tau_q)=-k\vec{\nabla}T(\vec{r},t+\tau_t)
	 	\label{ConstitutiveDPL}
	 \end{equation}
 
In Eq. \ref{ConstitutiveDPL}, two phase lags of $\tau_T$ and $\tau_q$ represent, respectively, the phase lag of the heat flux and the temperature gradient. In comparison to the existing microscopic models, the lagging behavior is used in the place of electron/phonon coupling in metals \cite{Qiu1992}, umklapp and normal phonon scattering \cite{Guyer1966}, and other relaxation of internal energy \cite{Gurtin1968}. Both thermal phase-lag values are determined experimentally \cite{Roetzel2003} and theoretically \cite{Basirat2009}.

Meanwhile, significantly fewer researches were devoted to the spatial attitude of the non-Fourier heat transfer. In principle, the quasi-ballistic heat flux in a given location and at a given time intrinsically depends on the temperature gradient in other places, in addition to the earlier times \cite{Mahan1988}. In other words, the constitutive law is no longer localized in non-diffusive transport regime, but instead possesses \textquote{memory} in space alongside the time memory. From a microscopic point of view, a convolution kernel $\kappa^{*}$, which is called the nonlocal thermal conductivity, fully contains the spatio-temporal memory of the heat flux concerning the temperature gradient \cite{Vermeersch2014}. This parameter appears in the following postulated constitutive law, or the so-called flux-gradient relation (FGR), which holds when the conventional Fourier law no longer does hold:

 \begin{equation}
 	q(x,t)=-\int_{0}^{t} dt'\int_{-\infty}^{\infty} dx' \kappa^{*}(x-x',t-t')\frac{\partial T}{\partial x}(x',t')
 	 	\label{FGR}	
 \end{equation}

Here, the convolution kernel $\kappa^{*}$, is the \textquote{nonlocal thermal conductivity kernel} of the medium. $\kappa^{*}$ proceeds as an intrinsic material characteristic of the thermal conducting medium. Under the relaxation time approximation, the Boltzmann transport equation conventionally follows the postulated constitutive law. After Fourier and Laplace transformations and inserting the analytical single pulse response of the RTA-BTE, one yields a generic expression regarding the underlying microscopic phonon dynamics  \cite{Vermeersch2014}. Once the kernel is obtained, the nonlocal heat transport is maintained. For the thermal transport through the length and time scales that are relatively larger than the phonon mean free paths and relaxation times, $\kappa^*$ reaches the constant value and, consequently, the FGR depicted by Eq. \ref{FGR}, becomes localized and reduces to Fourier's law.

On the other side, from a macroscopic viewpoint, the equivalency between the spatial nonlocal behavior and the lagging response in time, had been confirmed \cite{Tzo95a,DYTzou1997}. However, due to the absence of a constitutive model supporting, both were not considered concurrently. In 2010, Tzou and Guo, extended the concept of thermal lagging, in time, to the nonlocal response in space \cite{DYTzou2010}. The study was endorsed under the phonon scattering, the Guyer–Krumhansl model, and the thermomass model, which expresses the same nonlocal behavior as the first- and second-order effects. At first, nonlocal behavior with thermal lagging was motivated by the famous Guyer–Krumhansl model \cite{Guyer1966}. The model described via the following equation, includes the parameters which characterize the heat wave transition through the dielectric crystals:

\begin{equation}
	\label{Guyer}
	\textbf{q}+\tau_R\frac{\partial\textbf{q}}{\partial t}+(\frac{c^2\tau_R c^2}{3})\nabla T=\frac{\tau_R \tau_N c^2}{5}[\nabla^2\textbf{q}+2\nabla(\nabla.\textbf{q})].
	\end{equation}

The above constitutive equation for heat transport in phonon systems, introduces two relaxation times, $\tau_R$, and $\tau_N$, which, respectively, describe the umklapp (momentum-nonconserving), and normal (momentum-nonconserving) processes in a phonon framework. The relaxation times are related to two phase lags of DPL via $\tau_q$=$\tau_R$ and $\tau_T$=9$\tau_N$/5 \cite{DYTzou1997}. The $\tau_R \tau_N c^2$/5 on the right-side of the Eq. \ref{Guyer}, has a length squared dimension. The lengths $\tau_R c$ and $\tau_N c$ are proportional to the phonon path traveled during $\tau_R$ and $\tau_N$, and so $\tau_R \tau_N c^2$/5$\approx l^2$. The first-order effect in "$l$" does not appear in Guyer–Krumhansl equation. It is necessary to conform this spatial accomplishment together with the temporal effect to clear out the heat transport in phonon systems. 

Besides, the thermomass (TM) model \cite{Cao2007} is also another framework that supports the presence of the nonlocal behavior together with thermal lagging while investigating heat transport. Through the TM model, where a finite mass calculated from Einstein's mass-energy relation is attributed to phonons, the energy and constitutive equation in heat transport are derived, subsequently, from the continuity and the momentum equation. In consequence, the continuity equation reduces to the energy equation in the phonon gas:

\begin{equation}
	\label{energy-tm}
	\frac{\partial \rho_h}{\partial t}+\frac{\partial}{\partial x}(\rho u_h)=0 \ \implies -\frac{\partial q}{\partial x}=C\frac{\partial T}{\partial t}.
	\end{equation}

Also, the constitutive equation in one-dimensional reads out:

\begin{multline}
	\label{thermomass}
	\hspace{-0.5cm}
\rho_{h}\overbrace{(\frac{\partial u_h}{\partial t}+u_h\frac{\partial u_h}{\partial x})}^{\frac{D u_h}{Dt}}+\frac{\partial P_h}{\partial x}+f_h=0 \; \implies  \; \tau_{TM}\frac{\partial q}{\partial t}-\\ \Lambda C \frac{\partial T}{\partial t}+\Lambda \frac{\partial q}{\partial x}
-M^2k\frac{\partial T}{\partial x}+k \frac{\partial T}{\partial x}+q=0.
\end{multline}

Here, $\tau_{TM}$, the lagging time in the thermomass model, is about two orders of magnitude larger than the heat flux phase lag in the CV model. Also, $l=qk_p/[2\gamma C(CT)^2]$ and M are, respectively, the length parameter, and the drift velocity thermal Mach number with respect to the thermal wave speed. When the heat flux is replaced from Eq. \ref{energy-tm} to Eq. \ref{thermomass}, one obtains an energy equation that contains only the temperature:

\begin{equation}
	\frac{k(1-M^2)}{C\tau_{TM}}\frac{\partial^2T}{\partial x^2}=\frac{1}{\tau_{TM}}\frac{\partial T}{\partial t}+\frac{2l}{\tau_{TM}}\frac{\partial^2 T}{\partial x \partial t}+\frac{\partial^2 T}{\partial t^2}.
	\label{Thermomassmodel}
\end{equation}

The parameters involved in the above equations are $\rho_h$, $u_h$, $p_h$, and $f_h$, which are, respectively, defined as the phonon gas thermomass density, the drift velocity, the pressure of the phonon gas, and the resistance force per unit volume. Although the wave behavior is understood by the second order time derivative, $\partial^2 T/\partial t^2$, Eq. \ref{Thermomassmodel} also consists of a mixed-derivative term, $\partial^2 T/\partial x \partial t$. This part is new to the field of microscale heat transfer. The standard DPL model, considering only the phase lags, does contain similar mixed-derivative terms, which are even orders with respect to $x$. As it is seen, the length parameter, $l$, in Eq. \ref{Thermomassmodel}, is proportional to the heat flux, $q$. 

The presence of the length parameter in Eqs. \ref{Guyer} and \ref{Thermomassmodel} is explanatory for including the nonlocal attitude in space, as well as the lagging time in the CV model. Such nonlocality and lagging are implied in Fourier's law through the succeeding statement:

\begin{equation}
	\label{nonlocal-cv}
	q(x+\lambda_q,t+\tau_q)=-K\frac{\partial T}{\partial x}(x,t).
	\end{equation}

The equation supports that the temperature gradient through the material volume on position r and at time t, is proportional to the heat flux vector flowing across the other volume element positioned at ($r + L$) at a later time (t + $\tau_q$). The parameters, $K$=$k(1-M^2)$, $\tau_q$=$\tau_{TM}$, and $\lambda_q$=2$l$, ascertain that the nonlocal model is equivalent to the TM model. In the next step, the nonlocality is implied to more accurate DPL model \cite{Ghazanfarian2015,Shomali2021}. Consequently, the Nonlocal DPL equation becomes:

\begin{equation}
	\label{nonlocal-dpl}
	q(x+\lambda_q,t+\tau_q)=-K\frac{\partial T}{\partial x}(x,t+\tau_T).
\end{equation}

On the other hand, the heat transport study of the transistors is particularly significant as their reliability is determined via the obtained maximum temperature during the operation \cite{Samian2013,Samian2014,Moore2014}. When the temperature and heat flux distributions are calculated, we get information about how much temperature increase, each position inside the single transistor or on a whole die, experiences \cite{Shomali2012,Shomali20152,Shomali2016,Shomali2017,2Shomali2017,Shomali2022,Shomali2023}. This temperature augmentation in a MOSFET occurs due to the response to the existent self-heating \cite{EPop2005,EPop2006,EPop2010,Gong2015}. Consequently, predicting the temperature and heat flux distribution preciously forasmuch as the reliability is determined can help the engineers to design MOSFETs with optimal thermal conditions. Here, the nonlocal DPL model has been utilized for the thermal investigation of the 1-D MOSFETs. Newly nonlocality, phase lagging and, temperature jump coefficients are derived. Explicitly, the obtained scaled parameters are found to be almost different from that of the localized standard DPL model. The following study confirms that the implication of the nonlocal effects notably makes the simulation yield a more accurate results \cite{Ghazanfarian2009}. To be more precise, although for low Knudsen number devices, the phenomenological DPL model brings out less meticulous results, taking nonlocality in space for DPL, leads to much more accurate temperature profiles and heat flux plots. In this paper, first, in Sec. \ref{Sec.2}, the studied geometry and the relevant boundary conditions are given. Then in Sec. \ref{Sec.3}, the mathematical modeling for NDPL is developed. Section. \ref{Sec.4} is devoted to explaining about the numerical method. Finally, the results are pointed out in Sec. \ref{Sec.5} and, the paper is concluded in Sec. \ref{Sec.6}.

\section{Geometry and Boundary conditions}
\label{Sec.2}

\begin{figure}
 	\centering
 	\includegraphics[width=\columnwidth]{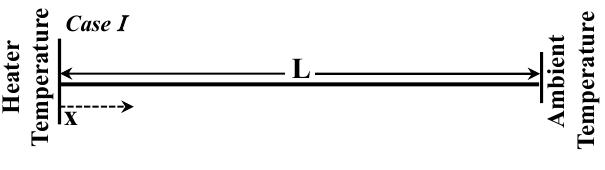}
 	\caption{\label{geo} Schematic geometry of the studied cases: A thin slab of silicon device (a) without and (b) with uniform heat generation zone.}
 \end{figure}

A one-dimensional MOSFET device is modeled in two cases. First, as shown in Fig. \ref{geo} (a), a thin slab of silicon transistor with a constant initial temperature of $T_0=300 \ K$ is studied. The system is the same as that of the one investigated by Ghazanfarian and Abbassi \cite{Ghazanfarian2009}. The temperature of the bottom boundary has abruptly increased. This results in the rearrangement of the temperature distribution in the slab. The heat transport in the thin film is studied considering the lagging responses and the nonlocality in space simultaneously. During the calculation, the applied temperature at the left boundary is kept constant at room temperature. The size of the transistor, L, is selected as the characteristic length of the simulation and manages the value of the Knudsen number, defined as the ratio of the phonon mean free path to the characteristic length, $Kn=\frac{\lambda}{L}$.

\section{Mathematical Modeling}
\label{Sec.3}

In 1995, Tzou proposed a non-Fourier approximation for heat conduction, in which the heat flux vector at a point in the material at time $t+\tau_q$ corresponds to the temperature gradient at the same point at time $t+\tau_t$:

\begin{equation}
	\vec{q}(\vec{r},t+\tau_q)=-k\vec{\nabla}T(\vec{r},t+\tau_t).
	\label{ConstitutiveDPL2}
\end{equation}

Where $\tau_q$ and $\tau_t$, respectively, represent the heat flux and the temperature gradient phase lag, both positive and inherent characteristics of the material. The thermal phase-lag values, like any other thermal characteristics for engineering materials, can be determined theoretically or experimentally under various conditions. Tzou \cite{DYTzou1997} has tabulated some analytical findings for phase lags. Also, Basirat \emph{et al}. \cite{Basirat2006} have reported values for the phase lags of some metal films. In the present research, the DPL model will be called the standard DPL. In 2010, Tzou generalized the DPL model by introducing the new term measuring the nonlocality of the heat flux vector. Consequently, the novelly proposed equation is:

\begin{equation}
	q(x+\lambda_q,t+\tau_q)=-K\frac{\partial T}{\partial x}(x,t+\tau_t).
	\label{nonlocalityTM}
\end{equation}

The heat flux, $\tau_q$, and temperature, ($\tau_t$), phase lags, are considered small in comparison to the time scale, and hence only the first order Taylor expansion of the DPL equation relative to $\tau_q$ and $\tau_t$ are kept. The situation is the same while considering nonlocality for the heat flux. The correlating length ($\lambda_q$) is  also modest compared to the space dimension. Accordingly, the Eq. \ref{nonlocalityTM} containing the first-order effects of $\tau_q$, $\tau_T$, and $\lambda_q$ then yields:

\begin{equation}
	q(x,t)+\lambda_q\frac{\partial q(x,t)}{\partial x}+\tau_q\frac{\partial q(x,t)}{\partial t}=
	-K\frac{\partial T}{\partial x}(x,t)-K\tau_T\frac{\partial^2 T}{\partial x \partial t}(x,t).
	\label{firstorderexpansionDPl}
\end{equation}

Moreover, the conventional energy equation is,

\begin{equation}
-\frac{\partial q}{\partial x}=C\frac{\partial T}{\partial t}.
	\end{equation}

Taking the derivative of the Eq. \ref{firstorderexpansionDPl} relative to the position, one can eliminate the heat flux terms in Eq. \ref{firstorderexpansionDPl} using the energy equation,

\begin{equation}
	(\frac{K}{C\tau_q})\frac{\partial^2 T}{\partial x^2}+(\frac{K\tau_T}{C\tau_q})\frac{\partial^3 T}{\partial x^2 \partial t}=\frac{1}{\tau_q}\frac{\partial T}{\partial t}+(\frac{\lambda_q}{\tau_q})\frac{\partial^2 T}{\partial x \partial t}+\frac{\partial^2 T}{\partial t^2}.
	\label{energynonlocalityDPL}
\end{equation}

In the next step, to obtain the normalized equations, the following non-dimensional parameters are defined:

\begin{equation}
	\theta=\frac{T-T_0}{T_W-T_0},  \ t^{*}=\frac{t}{\tau_q},\  B=\frac{\tau_t}{\tau_q}, \ \eta=\frac{x}{L},\ Kn=\frac{\lambda}{L}, \ \gamma=\frac{\lambda_q}{L}.
\end{equation}

Here, $\lambda$ and $L$, respectively, the phonon mean free path and the length of the system, determine the Knudsen number, $Kn$. Using the non-dimensional parameters, the Eq. \ref{energynonlocalityDPL} is rewritten this way:

\begin{equation}
	\frac{\partial\theta}{\partial t^*}+\gamma\frac{\partial^2 \theta}{\partial \eta \partial t^*}+\frac{\partial^2 \theta}{\partial t^{*^2}}=\frac{Kn^2}{3}\frac{\partial^2 \theta}{\partial \eta^2}+B\frac{Kn^2}{3}\frac{\partial^3 \theta}{\partial \eta^2 \partial t^*}.
	\label{nondimensionalnonlocalDPL}	
\end{equation}

Previously in 2002, Chen \cite{Chen2002} simulated the one-dimensional silicon transistor using the one-dimensional Ballistic-Diffusive Equations and the Boltzmann equation for the phonons. In the following, the accuracy of our results obtained utilizing NDPL, will be checked by comparison to Chen's findings. 

\subsection{Boundary conditions}
As the DPL model does not consider the boundary phonon scattering effects, enforcing no-jump boundary conditions leads to unsatisfactory results, especially near the boundaries \cite{Ghazanfarian2009}. Hence, here, the temperature jump boundary condition is used to fix the problem:

\begin{equation}
	\theta_s-\theta_w=-\alpha Kn(\frac{\partial \theta}{\partial \eta})_w.
	\label{boundrycondition}
\end{equation}

Where, $\theta_s$ and $\theta_w$ are, subsequently, the wall's jump temperature and the boundary temperature. Also, $\alpha$, is the coefficient which will be adjusted to satisfy the boundary condition. Also, two unknown parameters of $B$ and $\gamma$ alongside $\alpha$, appearing in Eq. \ref{nondimensionalnonlocalDPL}, are determined such that the results of the NDPL model becomes compatible with the solution of the phonon Boltzmann equation (PBE). Interestingly, it is obtained that the values of these parameters depend on the Knudsen number and time. Notably, according to the available data, the following values for $\alpha$, $B$, and $\gamma$ are attained:

\begin{equation}
	\alpha=\begin{cases}
		0.35t, & t \le 0.1\\
		0.27t, &  0.1 \le t \le 1 \\
		0.55, &  t>1 \ \& \ Kn>1 \\
		0.65-0.1\times(Kn), &  t>1 \ \& \ Kn<1. \\
	\end{cases}
	\label{a}
\end{equation}
\begin{equation}
	B=\begin{cases}
		0.45t, &  t \le 0.1,\\
		0.35t, & 0.1 \le t \le 1, \\
		0, & t > 1.
	\end{cases}
	\label{B}	
\end{equation}
\begin{equation}
	\gamma=0.35\times (Kn);  for \ all \ t.
	\label{g}   
\end{equation}	

Using the above-obtained constants, results in temperature and heat flux profiles that are very close to what is calculated from the Boltzmann equation and also the ballistic-diffusive equations. It should be mentioned that, here, the thermal properties such as thermal conductivity, sound velocity, and specific heat are considered constant.

%---------------------------
\section{Numerical solution}
\label{Sec.4}

The Eq.~\ref{energynonlocalityDPL} is solved using a completely implicit first- and second-order finite difference method. Discretizing of all derivatives is central, where a stable and convergent three-level finite difference scheme is utilized \cite{Dai2004}. Accordingly, the discretizations are written as below:

\begin{equation}
	\frac{\partial \theta}{\partial t^*}=\frac{1}{2\Delta t}[T^{n+1}_i - T^{n-1}_i],
\end{equation}
\begin{equation}
	\frac{\partial^2 \theta}{\partial \eta \partial t^*}=\frac{1}{4\Delta x \Delta t}[(T^{n+1}_{i+1}-T^{n+1}_{i-1})-(T^{n-1}_{i+1}-T^{n-1}_{i-1})],
\end{equation}
\begin{equation}
	\frac{\partial^2 \theta}{\partial t^{*^2}}=\frac{1}{\Delta t^2}[T^{n+1}_{i}-2T^n_{i}+T^{n-1}_{i}],
\end{equation}
\begin{equation}
	\frac{\partial^2 \theta}{\partial \eta^2}=\frac{1}{\Delta x^2}[T^{n+1}_{i+1}-2T^{n+1}_i+T^{n+1}_{i-1}],
\end{equation}
and,
\begin{equation}
	\frac{\partial^3 \theta}{\partial \eta^2 \partial t^*}=\frac{1}{2\Delta x^2\Delta t}[(T^{n+1}_{i+1}-2T^{n+1}_{i}+T^{n+1}_{i-1})-(T^{n-1}_{i+1}-2T^{n-1}_{i}+T^{n-1}_{i-1})].
\end{equation}

Here, a weighted average is implied for stability and convergence. The convergence of the numerical method is bettered by the Knudsen number reduction. It is to say that for high Knudsen numbers, the solution explicitly depends on the marching step size.

%-----------------------------------------------------------------------------
\section{Results and Discussions}
\label{Sec.5}
\begin{figure*}[h!]
	\centering
	\includegraphics[width=2\columnwidth]{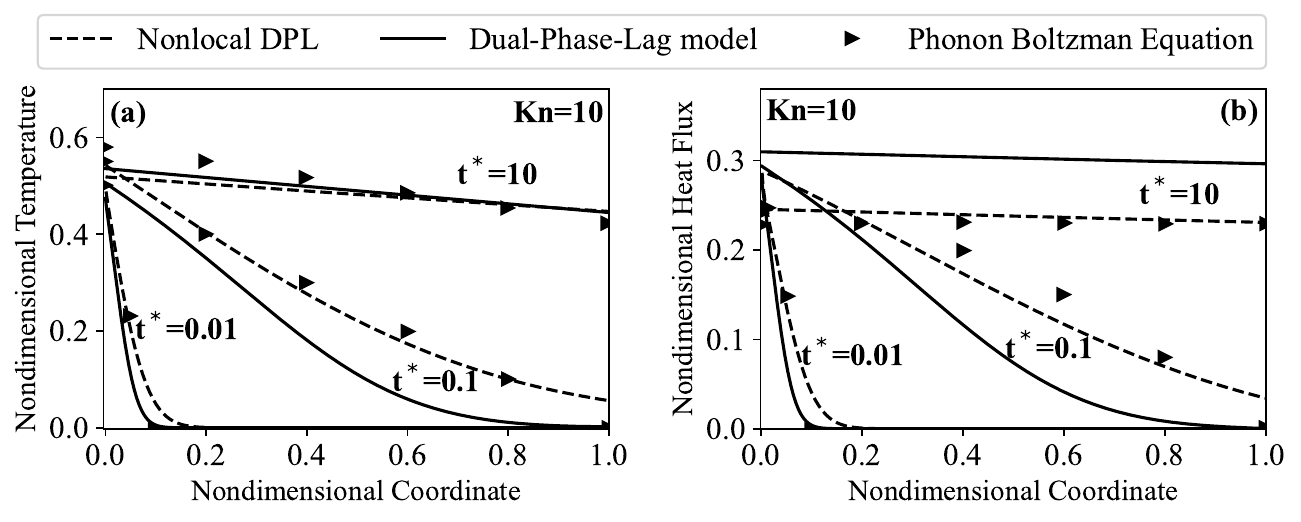}
	\caption{\label{kn10} The temperature profile calculated from NDPL, DPL, and PBE when Kn=0.1}
\end{figure*}

\begin{figure*}[h!]
	\centering
	\includegraphics[width=2\columnwidth]{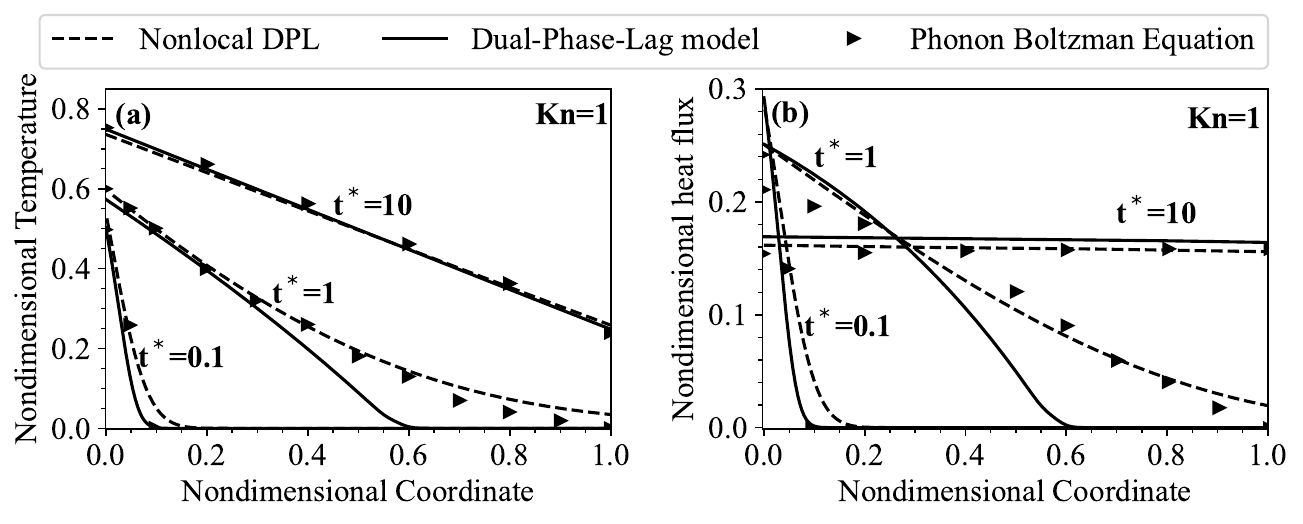}
	\caption{\label{kn1} The same as Fig. \ref{kn10} but for Kn=1.}
\end{figure*}

\begin{figure*}[h!]
	\centering
	\includegraphics[width=2\columnwidth]{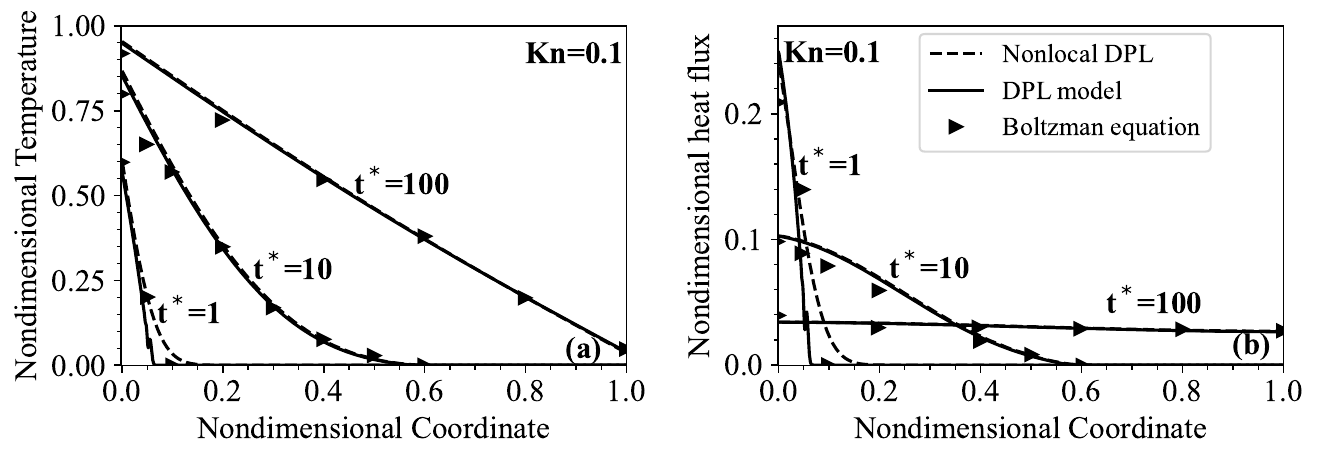}
	\caption{\label{kn01} Comparison of the temperature distribution obtained from the Boltzmann Equation, right triangle, the nonlocal behavior DPL model, dashed line, and the standard DPL model for different instantaneous times and Kn=0.1.}
\end{figure*}

The results for modeling the heat transport in 1-D MOSFET using the numerical simulation of the NDPL model, are manifested in Figs. \ref{kn10}-\ref{kn01}. Specifically, the comparison of the results for the non-dimensional temperature profile and the heat flux distribution, obtained from the nonlocal DPL, standard DPL, and PBE are presented. It should be made clear that verification of the output profiles is performed using the results achieved from PBE. Along the present NDPL simulation, the profiles for standard DPL modeling are also reproduced by the authors. 

In the work \cite{Ghazanfarian2009}, the temperature and heat flux profiles for 1-D MOSFETs with different Knudsen numbers, at various times, have been presented using the standard DPL model. While the Knudsen number grows, the deviation of the results from the precious profiles of the PBE increases. The present study will try to solve the large Knudsen number issue, using the nonlocal heat transport concept. More precisely, the nonlocality of heat flux is considered in the system by introducing the non-dimensional correlating length of $\gamma$=3.5. This almost large $\gamma$ confirms that taking into consideration the nonlocality is very crucial for large Knudsen number nano-structures. Trying to clarify, the better estimate of the results obtained from the NDPL as compared to that of the standard DPL model is shown in Fig. \ref{kn10} for the 1-D transistor with the Knudsen number of 10. Specifically, Fig. \ref{kn10} (a) presents the temperature profile for the case with Kn=10. As the figure suggests, the behavior of the temperature distributions obtained from NDPL for the scaled times of 0.01 and 0.1, are very much like the PBE solutions. It is obvious that the compatibility of the NDPL-obtained temperature distribution with PBE results is always better than those attained from DPL. This consistency is so that the NDPL calculated temperature behavior, completely fits the PBE result for scaled coordinates below x$^*$=0.1 and 0.8, respectively, for t$^*$=0.01 and 0.1. When t$^*$=0.01, both the size and time are very low, and the non-Fourier behavior of the system is very significant. For t$^*$=0.1, the system still experiences non-Fourier conditions, but it is a little less in comparison to that of the t$^*$=0.01. As evident in Fig. \ref{kn10} (a), considering nonlocality in the DPL model, handles non-Fourier behavior in nearly all positions such that the temperature distribution almost equals the PBE result. This is true while the result for big times like t$^*$=10, does not vary considerably with adding nonlocality characteristics. This department is justified as although the size of the system is petite, but at significant times it reaches a steady state, and consequently the non-Fourier behavior is adjusted. 

The heat flux versus position is demonstrated in Fig. \ref{kn10} (b). The very better consistency of the NDPL modeling compared to standard DPL, with the PBE data, is further seen. When t$^*$=0.1, for positions below x$^*$=0.01, similar to the temperature profile, the heat flux matches well with PBE results. Moreover, when t$^*$=0.1 and for x$^*$>0.2, the NDPL obtained heat flux profile displays considerable improvement over what the standard DPL calculates. More concretely, for large Knudsen numbers, the heat flux distribution obtained from DPL modeling behaves worse than the computed temperature profile in a way that gives a much inferior approximation of the exact outcomes of PBE. Interestingly, adding nonlocality to the DPL model, solves the problem and leads to the more correct solutions. For larger times, the nonlocality, and lagging behavior, loses its importance and $\gamma$ and B become zero. This is expected as the transistors reach the steady state and the role of non-Fourier heat transfer becomes much less prominent. As Fig. \ref{kn10} (b) suggests, $\alpha$, the temperature jump coefficient, is the only responsible parameter for achieving appropriate temperature and heat flux profiles. In this research, for large ts, the Knudsen number dependent $\alpha$ is determined such that it severely rectifies the very different standard DPL obtained heat flux profile.

Figure. \ref{kn1} shows the state of the 1-D transistor with the Knudsen number of 1. This Kn is neither high nor low. The trend for temperature and heat flux behavior is almost similar to that for Kn=10. Here we analyze the results in more detail. For lower Knudsen numbers, it is anticipated that the non-Fourier attitude and, consequently the nonlocality effect fades. The non-dimensional nonlocality presenter, $\gamma$, is 0.35, which is one-tenth the one for Kn=10. As Fig. \ref{kn1} (a) confirms, this almost low $\gamma$, is essential in getting results that are more congruent to the PBE solutions. Also, as non-Fourier behavior has become a little less important in comparison to the case with Kn=10, considering $\gamma$ makes the NDPL calculated profiles to be very accurately fitted to the available precise data \cite{Chen2002}. In other words, almost the complete deviation of the standard DPL results from PBE outputs', is filled via the nonlocality contribution. For instance, when t$^*$=0.1, for a broader range of position, say for x$^*$<1.5 and x$^*$>2, the NDPL temperature profile finding, provides the results found out solving the PBE. The consistency is more noticeable when we are dealing with t$^*$=1. For this time, compatibility and fitness exist for almost all positions. In consequence, taking into account the temperature profile obtained from standard DPL, it becomes pretty clear how our new model is doing much better. Although, for Kn=10, the NDPL distributions were close to the PBE solutions, the better agreeableness at Kn=1.0, is attributed to the slightly weaker non-Fourier behavior, which is practically covered by the simultaneous consideration of nonlocality and phase lagging. When a long time passes, the situation becomes like in nano-structures with larger Kn numbers. In similarity, the parameters $\gamma$ and $B$ reach zero, and the effect of nonlocality and phase lagging disappears. 

On the other side, in Fig. \ref{kn1} (b), the heat flux demeanor versus position is illustrated. Same as the case with Kn=10, here also, the NDPL modeling assures the results very close to the available atomistic data. As it is viewed, when t$^*$=0.1, like what happens to temperature profile, the NDPL calculated heat flux, is consistent with available data for x$^*$<0.1. Only as it is the case for Kn=10, the heat flux at x$^*$=0, is a bit larger than that of PBE. Moreover, heat flux fits preciously for x$^*$>0.2, and well for x$^*$<0.2, with the existing data. One can conclude that although NDPL predicts the heat flux correctly, a slight deviation exists for meager non-dimensional positions. It is important to note that the accuracy of the heat flux profile obtained from standard DPL is reported to be always less than the DPL calculated temperature distribution, while the accuracy is defined in closeness to the PBE results. Intriguingly, the NDPL resolves the heat flux inaccuracy and gives a highly meticulous distribution. Also, for a larger scaled time of t$^*$=10, when $\gamma$ and B are zero, the newly defined Kn dependent $\alpha$, reproduces the PBE solution.

At last, the temperature and heat flux distributions are studied for low Knudsen numbers. The results are shown in Fig. \ref{kn01}. As low Knudsen number systems have larger sizes, one anticipates less anomalous heat transfer inside. This does not mean that the effect of nonlocality on heat flux disappears. Even though the non-dimensional nonlocality of the heat flux for Kn=0.1, is $\gamma$=0.035, which is two orders of magnitude less than $\gamma$=3.5 for Kn=10, it is crucial to consider such nonlocality to get more precise plots. As Fig. \ref{kn01}(a) suggests for the lower scaled time of t$^*$=1, where the Non-Fourier attitude is bold compared with larger times of t$*$=10 and t$^*$=100, the NDPL presents a temperature profile with signifiently much better consistency with the PBE relative to the standard DPL. On the other hand, the small Knudsen numbers and large scaled times, say t$^*$>1, change the conditions in favor of removing the nonlocality and phase lag effects. So, while $\gamma$ and B become zero, the MOSFET reaches the steady state, and consequently, both the DPL and NDPL calculate distributions nearly the same as the PBE findings.

The imprecision in calculating the heat flux for low Kn number and smaller times using the DPL model is also worked out, implying the newly proposed NDPL. Just as it is evident in Fig. \ref{kn01}, for kn=0.1 and t$^*$=1, the NDPL obtained heat flux presents a remarkably improved condition such that at x$^*$=0 is closer to the PBE result and also fits it for a much wider range of positions. For larger times, as previously mentioned for temperature distribution, both DPL and NDPL work well while the non-Fourier effect is adjusted.

%-----------------------------------------------------------------------------------
\section{Conclusion}
\label{Sec.6}

The current research tries to establish a framework for micro/nanoscale heat transport investigation that produces precious results while concurrently having low computational costs. To achieve this aim, the 1-D nonlocal DPL model with the possibility of generalization to the higher dimensions is developed. This method considers the nonlocality of the heat flux along the phase lagging, while the temperature jump is also enforced. In particular, a non-dimensional parameter $\gamma$, in addition to the $\alpha$, the temperature jump coefficient, and $B$, the phase lagging ratio, both defined in DPL, is introduced to determine the strength of the nonlocality. Then, the formulated computational procedure is applied to a 1-D silicon MOSFET. The parameters $\gamma$, $\alpha$, and $B$ are newly evaluated by verifying of the obtained results with PBE-calculated available data. The presence of the Knudsen-dependent nonlocality coefficient, $\gamma$, is found to be crucial for obtaining accurate temperature and heat flux distribution. In further detail, as expected, for the large Knudsen number, the nonlocality at heat flux is very distinguished. As an instance, when Kn is 10, the $\gamma$ parameter becomes 3.5, which is one hundred times the $\gamma$ value for Kn=0.1. This does not mean that nonlocality is not essential for low Knudsen numbers. Taking into account the small $\gamma$ also makes the results very close to the PBE available data, especially for earlier times. In addition, introducing nonlocality to the DPL model, resolves the problem of inaccurate heat flux obtained from DPL and corrects it to remarkably acceptable value. Finally, when the heat transport inside a MOSFET is predicted well, it is more feasible to propose new transistors supporting the thermal management solutions.
%--------------------------------------------------------------------


\begin{thebibliography} {10}
	
    \bibitem{Mahajan2002}
	R. Mahajan, R. Nair, V. Wakharkar, J. Swan, J. Tang, G. Vandentop, Emerging directions for packaging technologies, Intel Technol. J. 6 (2002) 2.
	
	\bibitem{Moghadam2014}
	M. Moghaddam, J. Ghazanfarian, A. Abbassi, Implementation of DPL-DD model for the simulation of nanoscale MOS devices, IEEE Transactions on Electron Devices, 61(9), 3131-3138, 2014.	
		
	\bibitem{Minnich2012}
	A.J. Minnich, Determining phonon mean free paths from observations of quasiballistic thermal transport, Physical review letters, 109(20), 205901, 2012.

	\bibitem{Shomali2019}
	J. Ghazanfarian, Z. Shomali, S. Xiong, 21st Century Nanoscience–A Handbook: Nanophysics Sourcebook (Volume One), Sattler, K. D. (Ed.), Chapter 4, CRC Press, 2019.

	\bibitem{Chiu2005}
	Y.  H. Chiu, V. V. Deshpande, H. C. Postma, C. N. Lau, C. Miko, L. Forro, and M. Bockrath, Ballistic phonon thermal transport in multiwalled carbon nanotubes, Physical review letters, 95(22), 226101, 2005.

	\bibitem{Alvarez2007}
	F. X. Alvarez, and D. Jou, Memory and nonlocal effects in heat transport: From diffusive to ballistic regimes, Applied physics letters, 90(8), 083109, 2007.
	
	\bibitem{Shomali2018}
	Z. Shomali, R. Asgari, Effects of low-dimensional material channels on energy consumption of nano-devices, International Communications in Heat and Mass Transfer, 94, 77-84, 2018.
		
	\bibitem{Qiu1992}
	T. Q. Qiu and C. L. Tien,Short-pulse laser heating on metals, International Journal of Heat and Mass Transfer, 35, 719, 1992.
	
	\bibitem{Guyer1966}
	R. A. Guyer and J. A. Krumhansl, Solution of the linearized Boltzmann equation, Physical Review, 148, 766, 1966.
	
	\bibitem{Gurtin1968}
	M. E. Gurtin and A. G. Pipkin, A general theory of heat conduction with finite wave speed,  \newblock{ Archive for Rational Mechanics and Analysis}, 31,113, 1968.
	
	\bibitem{Roetzel2003}
	W. Roetzel, N. Putra, S. K. Das, \newblock{ International Journal of Thermal Science} 42 (2003) 541.
	
	\bibitem{Basirat2009}
    H. Basirat Tabrizi, S. Andarwa, \newblock{ International Communication in Heat and Mass Transfer} 36 (2009) 186.
		
	\bibitem{Mahan1988}
	G. D. Mahan, and F. Claro, Nonlocal theory of thermal conductivity. Physical Review B 38(3) (1988) 1963.
	
	\bibitem{Vermeersch2014}
	B. Vermeersch, and A. Shakouri, Nonlocality in microscale heat conduction, arXiv preprint arXiv:1412.6555 (2014).
		
	\bibitem{Tzo95a}
	Da~Yu Tzou, \newblock The generalized lagging response in small-scale and high-rate heating, \newblock{International Journal of Heat and Mass Transfer} 38(17) (1995) 3231.

	\bibitem{DYTzou1997}
	D.Y. Tzou, Macro- to Microscale Heat Transfer: The Lagging Behavior. Taylor \& Francis, Washington, D.C., USA, 1997.
	
	\bibitem{DYTzou2010}
	D. Y. Tzou, and Z. Y. Guo, Nonlocal behavior in thermal lagging, \newblock{ International Journal of Thermal Sciences} 49(7) (2010) 1133.
	
	\bibitem{Cao2007}
	B. Y. Cao, Z. Y. Guo, Equation of motion of a phonon gas and non-Fourier heat conduction,  \newblock{ J. Appl. Phys.} 102 (2007) 053503.
	
	\bibitem{Ghazanfarian2015}
	J. Ghazanfarian, Z. Shomali, A. Abbassi, Macro-to nanoscale heat and mass transfer: the lagging behavior, Int. J. Thermophys 36 (2015) 1416.
	
	\bibitem{Shomali2021}	
	Z. Shomali, R. Kovács, P. Ván, I.V. Kudinov, and J. Ghazanfarian, \newblock{Recent Progresses and Future Directions of Lagging Heat Models in Thermodynamics and Bioheat Transfer}, \newblock{Continuum Mechanics and Thermodynamics}, 34:637–679, 2022.
	
	\bibitem{Samian2013}
	R.S. Samian, A. Abbassi, J. Ghazanfarian, Thermal investigation of common 2d FETs and new generation of 3-d FETs using Boltzmann transport equation in nanoscale, Int. J. Mod. Phys. C 24 (2013) 1350064.

	\bibitem{Samian2014}
	R.S. Samian, A. Abbassi, J. Ghazanfarian, Transient conduction simulation of a nanoscale hotspot using finite volume lattice Boltzmann method, Int. J. Mod. Phys. C 25 (04) (2014) 1350103.
	
	\bibitem{Moore2014}
	A.L. Moore, L. Shi, Emerging challenges and materials for thermal management of electronics, Mater. Today 17 (4) (2014) 163.
	
	\bibitem{Shomali2012}
	J. Ghazanfarian and Z. Shomali, \newblock{Investigation of dual-phase-lag heat conduction model in a nanoscale metal-oxide-semiconductor field-effect transistor}, \newblock{ International Journal of Heat and Mass Transfer}, 55(21-22):6231-7, 2012.
		
	\bibitem{Shomali20152}
	Z. Shomali, J. Ghazanfarian, A. Abbassi, \newblock{Investigation of bulk/film temperature-dependent properties for highly non-linear DPL model in a nanoscale device: the case with high-k metal gate MOSFET}, \newblock{ Superlattices and Microstructures}, 83:699, 2015.
	
	\bibitem{Shomali2016}
	Z. Shomali, A. Abbassi, J. Ghazanfarian, Development of non-Fourier thermal attitude for three-dimensional and graphene-based MOS devices, Appl. Therm. Eng. 104 (2016) 616.
	
	\bibitem{Shomali2017}
	Z. Shomali, B. Pedar, J. Ghazanfarian, A. Abbassi, \newblock{Monte-Carlo Parallel Simulation of Phonon Transport for 3D Nano-Devices}, \newblock{ International Journal of Thermal Sciences}, 114:139-154, 2017.
	
	\bibitem{2Shomali2017}
	Z. Shomali, J. Ghazanfarian, A. Abbassi, 3-D Atomistic Investigation of Silicon MOSFETs, In Proceedings of CHT-17 ICHMT International Symposium on Advances in Computational Heat Transfer, ICHMT Digital Library Online, Begel House Inc., 2017.
		
	\bibitem{Shomali2022}
	M. H. Fotovvat and Z. Shomali, \newblock{A time-fractional dual-phase-lag framework to investigate transistors with TMTC channels (TiS$_3$, In$_4$Se$_3$) and size-dependent properties}, 168, 207304, 2022.
		
	\bibitem{Shomali2023}
	Z. Shomali, An investigation into the reliability of newly proposed MoSi$_2$N$_4$/WSi$_2$N$_4$ field effect transistor: A monte carlo study, arXiv preprint arXiv:2305.04327 (2023).
		
	\bibitem{EPop2005}
	E. Pop, R.W. Dutton, K.E. Goodson, Monte Carlo simulation of joule heating in bulk and strained silicon, Appl. Phys. Lett. 86 (2005) 082101.
	
	\bibitem{EPop2006}
	E. Pop, S. Sinha, K.E. Goodson, Heat generation and transport in nanometer-scale transistors, Proc. IEEE 94 (8) (2006) 1587.
	
	\bibitem{EPop2010}
	E. Pop, Energy dissipation and transport in nanoscale devices, Nano Res. 3 (3), (2010) 147.
	
	\bibitem{Gong2015}
	S. Gong, L. Chen, H. Feng, Z. Xie, F. Sun, Constructal optimization of cylindrical heat sources surrounded with a fin based on minimization of hot spot temperature, Int. Commun. Heat Mass 68 (2015) 1.
	
	\bibitem{Ghazanfarian2009}
	J. Ghazanfarian, A. Abbassi, \newblock{Effect of boundary phonon scattering on Dual-Phase-Lag model to simulate micro-and nanoscale heat conduction}, \newblock{ International Journal of Heat and Mass Transfer}, 52(15-16) (2009) 3706.
	
	\bibitem{Basirat2006}
	H. Basirat, J. Ghazanfarian, P. Forooghi, Implementation of dual-phase- lag model at different Knudsen numbers within slab heat transfer, in: Proceedings of International Conference on Modeling and Simulation (MS06), August, 2006, Konia, Turkey, pp. 895899.
	
	\bibitem{Chen2002}
	G. Chen, \newblock{Ballistic-diffusive equations for transient heat conduction from nano to macroscale}, \newblock{ ASME J. Heat Transfer} 124 (2001) 320–328.
		
	\bibitem{Dai2004}
	W. Dai, L. Shen, R. Nassar, and T. Zhu, A stable and convergent three-level finite difference scheme for solving a dual-phase-lagging heat transport equation in spherical coordinates, \newblock{Int. J. Heat Mass Transfer} 47 (2004) 1817.
	
\end{thebibliography}
\end{document}